\documentclass[twocolumn,amsmath,amssymb]{revtex4}

\usepackage{graphicx}
\usepackage{dcolumn}
\usepackage{bm}

\usepackage{color}

\def\msun{M_\odot}

\begin{document}

\title{The most massive objects in the Universe}

\author{Daniel E. Holz$^1$ and Saul Perlmutter$^{2,3}$}
\affiliation{%
$^1$Theoretical Division, Los Alamos National Laboratory, Los Alamos, NM 87545\\
$^2$Department of Physics, University of California Berkeley, Berkeley, CA
94720-7300\\
$^3$Lawrence Berkeley National Laboratory, Berkeley, CA 94720}


\begin{abstract}
We calculate the most massive object in the Universe, finding it to be a cluster
of galaxies with total mass $M_{200}=3.8\times10^{15}\,M_{\odot}$ at $z=0.22$,
with the $1\sigma$ marginalized regions being
$3.3\times10^{15}\,M_{\odot}<M_{200}<4.4\times10^{15}\,M_{\odot}$ and
$0.12<z<0.36$.  We restrict ourselves to self-gravitating bound objects, and
base our results on halo mass functions derived from N-body simulations.  Since
we consider the very highest mass objects, the number of candidates is expected
to be small, and therefore each candidate can be extensively observed and
characterized.  If objects are found with excessively large masses, or
insufficient objects are found near the maximum expected mass, this would be a
strong indication of the failure of $\Lambda$CDM. The expected range of the
highest masses is very sensitive to redshift, providing an additional
evolutionary probe of $\Lambda$CDM. We find that the three most massive clusters
in the recent SPT $178\,\mbox{deg}^2$ catalog match predictions, while XMMU
J2235.3--2557 is roughly $3\sigma$ inconsistent with $\Lambda$CDM.
We discuss Abell 2163 and Abell 370 as candidates for the most massive
cluster in the Universe, although uncertainties in their masses
preclude definitive comparisons with theory.
Our findings motivate further observations of the highest mass end 
of the mass function. Future surveys will explore larger volumes, and the
most massive object in 
the Universe may be identified within the next decade. The mass distribution of
the largest objects in the Universe is a potentially powerful test of
$\Lambda$CDM, probing non-Gaussianity and the behavior of
gravity on large scales.
\end{abstract}

\maketitle


\noindent{\em Introduction}---Our Universe has a finite observable volume, and therefore within our Universe
there is a unique most massive object. This object will be a supercluster of
galaxies. Theoretical studies of the growth of structure have now matured, and
the mass of the most massive objects can be robustly predicted to the level
of a few percent. Furthermore, we are in the midst of a revolution in our
ability to conduct volume-limited samples of high-mass clusters, with
Sunyaev-Zel'dovich (SZ) and X-ray surveys able to provide complete samples at mass
$>5\times10^{14}\,\msun$ out to $z>1$. The masses of the most massive clusters
in the Universe are therefore a robust prediction of $\Lambda$CDM models, as
well as a direct observable of our Universe.

The cluster mass function is already being utilized as a probe of cosmology, and
in particular, of the dark energy
equation-of-state~\cite{2001ApJ...560L.111H,2001ApJ...553..545H,2002PhRvL..88w1301W,2003ApJ...585..603M,2006PhRvD..74b3512K,2006astro.ph..9591A,2008MNRAS.385.2025C}.
What additional value is there in singling out the very tail end of the mass
function, representing the most massive clusters in the Universe, for special
treatment? First, we note that these systems are in many ways the easiest to
find, as they are among the largest and brightest objects. They thus avoid many
selection effects which might plague lower mass cuts. In addition, these systems
constitute a very small sample (ideally, just one compelling candidate), and it
is possible to devote significant observational resources to studying them. One
might imagine coupled S-Z, X-ray, and weak lensing measurements, and thus the
masses of these systems will be among the best constrained of any systems. The
mass-observable relation for clusters is an essential component in using the
cluster mass function to measure properties of the dark energy, and therefore
there is a tremendous amount of ongoing work to characterize the masses of these
objects~\cite{2003ApJ...585..603M,2005PhRvD..72d3006L,2005ApJ...623L..63M,2006ApJ...650..538N,2006ApJ...650..128K,2008ApJ...672...19R,2009arXiv0910.3668W}.
Finally, because we are probing far down the exponential tail of the mass
function, these objects offer an unusually powerful constraint. If the most
massive object is found to have too large a mass (or especially, as explained
below, too small a mass), this {\em single object} will provide a strong
indication of non-Gaussianity or modified gravity~\cite{1998ApJ...494..479C}. An
excellent example of this 
is the high-redshift cluster XMMU J2235.3--2557 (hereafter
XMM2235)~\cite{2005ApJ...623L..85M},
which has been argued to be a 
few sigma inconsistent with
$\Lambda$CDM~\cite{2009ApJ...704..672J,2009PhRvD..80l7302J,2010arXiv1003.0841S}.
A similar approach
based on strong lensing has been presented in~\cite{2009MNRAS.392..930O}, which
considers the distribution of the largest Einstein radii in the Universe as a
probe of $\Lambda$CDM.
Although much work has focused on using halo statistics as a
probe of cosmology, here we focus on using the high-mass tails of precision mass
functions to make explicit predictions for current and future observations.


A critical question in one's attempt to determine the most massive object is to
define precisely what is meant by ``object''. The largest structure in the
Universe detected to date is the Sloan Great
Wall~\cite{2005ApJ...624..463G}, but the identification of this wall as a unique
object is sensitive to a (completely arbitrary) density threshold.
For our purposes we
define an object as a gravitationally self-bound, virialized mass
aggregation. These objects have decoupled from the Hubble flow, and represent
large local matter overdensities. This definition has the convenience of
robustly identifying objects (both in theory and observation).

\medskip
\noindent{\em Mass function}---Recent years have shown tremendous progress in
characterizing the mass function
of dark matter halos in cosmological N-body simulations. We have now
established, to better than 5\%, the expected number density of dark matter
halos as a function of mass and
redshift~\cite{2006ApJ...646..881W, 2007MNRAS.374....2R,
2008ApJ...688..709T}.
In the simulations underlying these precise mass function
expressions, the halos at the high-mass end are resolved by millions of
particles,
lending particular confidence
and robustness to the mass function in this regime. The
simulations are pure dark matter, and neglect
the influence of baryons. At smaller scales baryons could play a major
role in the density profile of the dark matter halos, and could potentially
impact the mass function of the objects themselves. At the large scales being
considered in this paper, the effects of baryons are expected to be
negligible. This is particularly true as our interest is in the mass function,
and hence the number density of these halos, not their density
profiles.

An important issue is the process by which a dark matter halo is identified and
characterized in a dark matter simulation~\cite{2001A&A...367...27W}. There are
two dominant approaches: friends-of-friends (FOF) and spherical overdensity
(SO). FOF defines a halo by contours of constant density, while SO
defines halos by the overdensity (compared to the mean or critical
density) within a spherical region. It has been argued that the mass associated
with SO can be most closely tied to observations of
clusters~\cite{2008ApJ...688..709T}. On the other hand, using an FOF with a
linking length of 0.2 corresponds closely to contours of density 200 times the
background density, which from spherical collapse models is a natural proxy for
the virial mass.
Because of the steep exponential in the mass function, our results are
essentially independent of these differences (see Fig.~\ref{fig:fig3}).

\begin{figure}
\centering 
\includegraphics[width=0.98\columnwidth]{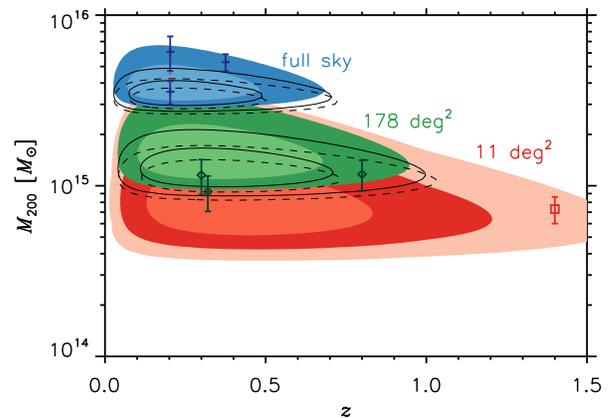}
\caption{\label{fig:fig2} Contour plot of the most massive object in the
Universe. Three sets of contours are provided, for three different surveys: full
sky, $178\,\mbox{deg}^2$ (corresponding to SPT), and $11\,\mbox{deg}^2$
(corresponding to XMM2235). The shaded contours represent the
$1\sigma$ and $2\sigma$ (and for the $11\,\mbox{deg}^2$ case, $3\sigma$)
regions of the most massive halo in a $\Lambda$CDM Universe. The solid
line contours are for the 2nd most massive halo, while the
dashed line contours are for the 3rd most massive halo. The (blue) plus signs
are Abell 2163 (double point) and Abell 370, the three (green) diamonds are the
three most massive clusters in the SPT $178\,\mbox{deg}^2$
survey, and the (red) square is XMM2235. Note that the mass values for
Abell 2163 span the predicted region, while Abell 370 is slightly high. The
SPT masses fit within their respective contours, while XMM2235
is well outside its $2\sigma$ contour. All masses are $M_{200}$: spherical
overdensity halos with $\Delta=200$ (measured with respect to $\rho_{\rm
matter}$). For data measured using different overdensities, we have
converted to the $M_{200}$ value which gives the equivalent probability.}
\end{figure}
\begin{figure}
\centering 
\includegraphics[width=\columnwidth]{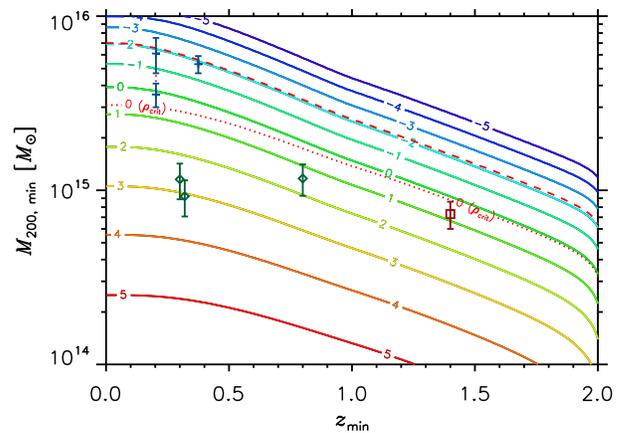}
\caption{\label{fig:fig3} Expected number of halos at redshift $\ge z_{\rm
min}$ with mass $\ge M_{200,\rm min}$, for a full sky survey. Each contour line
represents a value of $\log_{10}\left<{N}\right>$. For a survey with fraction, $f$, of the full
sky, the expected numbers of halos are diminished by the factor $f$. The dashed
(red) line shows the result for $\left<{N}\right>=0.01$ using the fit
from~\cite{2006ApJ...652...71W}, based on an FOF halo finder with $b=0.2$. It is
virtually indistinguishable from the corresponding SO ($\Delta=200$)
contour. The dotted (red) line represents the $\left<{N}\right>=1$ contour for a $\Delta=200c$
mass function, with overdensity compared to $\rho_{\rm crit}$, instead of the
average matter density, $\rho_{\rm matter}$. Note that this agrees with the
fiducial ``0'' line ($\Delta=200$) at high redshift, as the Universe becomes
matter dominated. The data points are the same as in
Fig.~\ref{fig:fig2}. Fitting forms for the curves in this figure are provided in
the text.}
\end{figure}

The halo mass function depends sensitively on cosmological parameters, including
$\Omega_m$, $\Omega_\Lambda$, and the equation-of-state of the dark energy.
For our purposes, one of the most important cosmological parameters is the
amplitude of the initial density fluctuations, characterized by $\sigma_8$, the
RMS variance of the linear density field, smoothed on scales of
$8\,\mbox{Mpc}$. Uncertainty in this quantity translates directly into
uncertainty in the amplitude of the mass function. We utilize the latest value
from {\it WMAP}, which provides a $\sim4\%$ measurement of
$\sigma_8$~\cite{2010arXiv1001.4538K}. For reference, a 5\% error on
$\sigma_8$ shifts the contours in Figure~\ref{fig:fig2} by less than $1\sigma$
in mass for a full-sky survey, and considerably less for smaller surveys.
Since the value of $\sigma_8$ is a major source of uncertainty in the use of the
cluster mass function to constrain cosmology, there is great interest in
improving its measurement. In addition, the mass function also depends
implicitly on the Hubble constant, $h$, which can be seen by expressing it in
units of $\mbox{\# of halos}/(\mbox{Mpc}/h)^3$ (observations naturally
measure volume in these units).
For simplicity we have explicitly put in
the {\it WMAP}\/7 value ($h=0.710$), but it is straightforward to re-express all
of our results explicitly in terms of $h$ (see Eqs.~\ref{eq:fit1}
and~\ref{eq:fit2}, and the text immediately beneath).

The mass function predicts the number density of massive dark matter halos in
the Universe. For the purposes of this paper we are also interested in
the scatter in this relation. At the high-mass end of the mass function, where
the number density satisfies roughly one per volume of interest, we assume that
the distribution of halos is given by Poisson statistics. This is valid 
as the largest objects are spatially independent on these scales
($>\mbox{Gpc}$), and are dominated by
shotnoise~\cite{2003ApJ...584..702H,2006PhRvD..73f7301H}.

We use the mass function presented in Tinker et
al.~\cite{2008ApJ...688..709T}, which gives the expected number density of dark
matter halos, $dn/dM$, in units of $\mbox{Mpc}/h$, where $h$ is the Hubble
constant and volume is measured in comoving $\mbox{Mpc}^3$. This mass function
describes the abundance of spherical-overdensity dark matter halos, and is
accurate to $\lesssim5\%$ over the redshift range of interest ($0<z<2$), and for
overdensity values (compared to the mean matter density at $z$) in the range
$200<\Delta<2300$. This mass function has been calibrated for
$M_{200}\lesssim4\times10^{15}\,\msun$, and therefore the extreme high-end of our
calculations relies on extrapolation.
In what follows we assume the {\it WMAP}\/7 cosmological
parameters, namely, $h=0.710$, $\Omega_m=0.264$, $\Omega_\Lambda=0.734$, and
$\sigma_8=0.801$~\cite{2010arXiv1001.4538K}.


\medskip
\noindent{\em The most massive object (Theory)}---We are interested in
determining the mass of the most massive object in our Universe.
We calculate the expected distribution of masses at the high mass end, assuming
Poisson statistics; the results are 
shown in Figure~\ref{fig:fig2}. The most massive object in the Universe is
expected to be found at $z=0.22$, with a mass
$M_{200}=3.8\times10^{15}\,M_{\odot}$. The marginalized $1\sigma$ range in
mass is $3.3\times10^{15}<M_{200}<4.4\times10^{15}$, while in redshift it is
$0.12<z<0.36$.
If the most massive object in the
Universe falls outside the range $2\times10^{15}\,\msun<M_{200}<10^{16}\,\msun$, we
can conclude with high confidence that either the initial 
conditions are non-Gaussian, or the growth of structure deviates from the
predictions of general relativity.

Figure~\ref{fig:fig2} includes contours of the 2nd and 3rd most massive halos in
the Universe. Going from the most massive to the 2nd most massive results in a
noticeable shift, demonstrating the power of just a few halos to constrain
cosmology. As we go further down (e.g., from the 2nd to the 3rd most massive),
the contours rapidly converge due to the exponential steepening in expected
number at lower mass. Note that the most massive halo occurs at low
redshift. Furthermore, the contours are not centered on the most likely point;
there is much larger scatter to high mass, with a sharp lower mass limit, due to
the exponential steepening. Note that these likelihoods are not
independent, since if the most massive object has an unusually low mass, it is
assured that the subsequent few most massive objects will also be unusually
low. We have performed Monte-Carlo studies which show that the
correlations are weak, however, and the distribution of separations is well
approximated by assuming the likelihoods are drawn independently.
Figure~\ref{fig:fig2} also shows the contours for
the 1st and 2nd most massive objects from the recent SPT $178\,\mbox{deg}^2$
survey~\cite{2010arXiv1003.0003V}, as well as the contours for the archival {\it
XMM-Newton}\/ survey which discovered XMM2235.



Figure~\ref{fig:fig3} shows contours of the expected number of halos greater
than a given mass, and found beyond a minimum redshift:
$\left<{N}\right>(>M_{200},>z)$. 
The contours are roughly linear in
the redshift range $0.2\lesssim z\lesssim2$, and are well approximated (to
better than 5\%) by the family of lines: $\log_{10}(M({\cal N},z))=a({\cal
N})+b({\cal N})z$ with
\begin{eqnarray}
\label{eq:fit1}
a({\cal N})&=&15.72-0.136{\cal N}-0.014{\cal N}^2-0.0012{\cal N}^3\\
b({\cal N})&=&-0.5375+0.00581{\cal N}+0.0024{\cal N}^2+0.00027{\cal N}^3, \nonumber
\nonumber
\end{eqnarray}
where ${\cal N}=\log_{10}\left<{N}\right>$. For the redshift
range $z<0.2$, the results are well represented by the values at
$z=0$, which are given (to better than $2\%$) by:
\begin{equation}
\log_{10}(M({\cal N}))=15.6-0.142{\cal N}-0.014{\cal N}^2.
\label{eq:fit2}
\end{equation}
These expressions can be utilized to calculate the expected number of objects
above a given minimum mass and redshift in the mass range
$10^{14}\,\msun<M_{200}<10^{16}\,\msun$ and redshift range
$0<z<2$, for any survey size. For a volume-limited sample, we are
interested in 
$\left<{N}\right>(>M_{200},<z)$. These contours start at 0 at $z=0$ (since there is
no volume), and rapidly rise to their maximum values, flattening by $z\sim0.2$
at the values given by Eq.~\ref{eq:fit2}.  Note that Eqs.~\ref{eq:fit1}
and~\ref{eq:fit2} assume the {\it WMAP}\/7 value of the Hubble constant,
$h=0.710$. To explicitly put in the $h$ dependence, $M_{200}$ and
$\left<N\right>$ can be rescaled by $(h/0.71)$ and $(0.71/h)^3$,
respectively.

\medskip
\noindent{\em The most massive object (Observations)}---The most
massive object in the Universe is likely to have already been detected by {\it
ROSAT}\/ (potentially even if it is behind the galactic
plane~\cite{2007ApJ...662..224K}). Reliably measuring the masses of candidate
{\it ROSAT}\/ 
sources remains challenging, however, and therefore the specific identity and
mass of the most massive object is unknown at present.  Perhaps the most
compelling candidate is Abell 2163 at $z=0.203$, which has an X-ray mass
measurement of $M_{500\rm
c}=3.4\pm0.8\times10^{15}\,\msun$~\cite{2009arXiv0909.3099M,2010Mantz_private}
(where ``500c'' indicates $\Delta$ with respect to $\rho_{\rm crit}$ rather than
$\rho_{\rm matter}$).  We expect 0.02 (0.002/0.2) clusters with at least this
mass and redshift in the entire Universe, where the numbers in parentheses are
the $1\sigma$ lower and upper bounds on $\left<{N}\right>$. An alternative, weak
lensing measurement of the mass yields a lower value of $M_{500\rm
c}=2.0\pm0.3\times10^{15}\,\msun$~\cite{2008A&A...487...55R}, which has
expectation 1.4 (0.5/4) (precisely agreeing with predictions). Furthermore,
\cite{2009ApJ...692.1033V} find an X-ray mass of $M_{500\rm
c}=2.3\pm0.07\times10^{15}\,\msun$, which agrees well with the
lensing value.
Abell 370 is another compelling
candidate, with a weak lensing mass of
$M_{vir}=2.93^{+0.36}_{-0.32}\times10^{15}\,h^{-1}\msun$ at
$z=0.375$~\cite{2008ApJ...685L...9B,2010MNRAS.402L..44R}, and an expectation of 0.02 (0.005/0.05).
These data points are shown in Figures~\ref{fig:fig2} and~\ref{fig:fig3}, 
where we have converted the masses to the $M_{200}$ values which give the
equivalent probabilities.

The figures also show the three most massive clusters from the SPT
$178\,\mbox{deg}^2$ survey~\cite{2010arXiv1003.0003V}, where we have added the
statistical and systematic errors in quadrature. For the most massive cluster
($M_{200}=(8.3\pm1.7)\times10^{14}\,\msun/h$ at $z=0.8$), we would expect 0.14
(0.04/0.5) clusters in the given sky area with a mass and redshift at least as
large. For the 2nd most massive cluster
($M_{200}=(8.2\pm1.9)\times10^{14}\,\msun/h$ at $z=0.3$), the expected number
goes up to 2 (0.8/6), while for the 3rd most massive
($M_{200}=(6.56\pm1.54)\times10^{14}\,\msun/h$ at $z=0.32$) we expect 5
(2/14). These masses are fully consistent with theory.

We also plot XMM2235, with a
mass of $M_{200\rm c}=(7.3\pm1.3)\times10^{14}\,\msun$ at
$z=1.4$~\cite{2009ApJ...704..672J}. This cluster was found in an
$11\,\mbox{deg}^2$ survey ($f=0.0003$). From Figure~\ref{fig:fig3} we would
expect to find a few thousand objects with at least this mass in the entire
Universe ($z>0$), and only 10 such objects at $z\ge1.4$ on the entire sky.  The
expected number of clusters in an $11\,\mbox{deg}^2$ survey, with this minimum
mass and redshift, is $1\times10^{-3}$ ($3\times10^{-4}/4\times10^{-3}$). A
conservative lower limit of $M_{324}=5\times10^{14}\,\msun$ is quoted
in~\cite{2009ApJ...704..672J}, which leads to an expectation of $6\times10^{-3}$
in the survey area (see also~\cite{2009PhRvD..80l7302J,2010arXiv1003.0841S}). From
Figure~\ref{fig:fig2} we see that XMM2235 
is a $3\sigma$ outlier. Alternatively, the cluster's true mass would have to be
reduced by $4\sigma$ to achieve $\left<N\right>=1$ (see
Figure~\ref{fig:fig3}). 
We note that these results are relatively insensitive to errors in the mass
determination; 15\% errors do not qualitatively alter our conclusions.

Current data argues for further exploration of the highest-mass end of the mass
function, both at low and high redshift. It would be particularly difficult,
theoretically, to account for excessively massive clusters at $z>1$, while
having agreement at lower redshift (e.g., non-Gaussianity would not suffice).
We expect to have dramatically improved complete high-redshift cluster surveys
with which to test $\Lambda$CDM in the near future, including the full SPT
survey ($2000\,\mbox{deg}^2$), the Dark Energy Survey ($5000\,\mbox{deg}^2$),
{\it Planck}\/ (all-sky), and eventually LSST ($20,000\,\mbox{deg}^2$).
In particular, {\it Planck}\/ is expected to provide a
relatively complete, all-sky survey of all massive clusters out to high
redshift in the near future~\cite{2003ApJ...597..650W}.
If the results from these cluster surveys disagree with the predictions
outlined above, the $\Lambda$CDM paradigm for the growth of structure will need
to be revisited.

\medskip
We acknowledge valuable discussions with Mark Bautz, Joanne Cohn, Bill
Holzapfel, Adam Mantz, Herman Marshall, Elena Pierpaoli, Paul Schechter, Jeremy
Tinker, Risa Wechsler, Martin White, and especially Jerry Jungman and Michael
Warren.



\bibliography{references}

\end{document}